\documentclass[a4paper,fleqn,usenatbib]{mnras}

\usepackage[T1]{fontenc}
\usepackage{ae,aecompl}

\usepackage{graphicx}	
\usepackage{amsmath}	
\usepackage{amssymb}	

\usepackage[utf8]{inputenc}
\usepackage{relsize}
\usepackage{wasysym}
\usepackage{soul}

\newcommand{\Gg}{\mbox{$G$}}
\newcommand{\GBP}{\mbox{$G_{\mathrm{BP}}$}}
\newcommand{\GRP}{\mbox{$G_{\mathrm{RP}}$}}
\newcommand{\Gabs}{\mbox{$G_{\mathrm{abs}}$}}
\newcommand{\Gmin}{\mbox{$G_{\mathrm{min}}$}}
\newcommand{\Gmax}{\mbox{$G_{\mathrm{max}}$}}

\newcommand\Gaia{\textit{Gaia}}
\newcommand\IMF{\textit{IMF}}
\newcommand\SFH{\textit{SFH}}
\newcommand\IFMR{\textit{IFMR}}
\newcommand\FMF{\textit{FMF}}
\newcommand\SEP{\textit{SEP}}

\title[Spectral Evolution on white dwarf from \textit{Gaia}]{Evidence of Spectral Evolution on the white dwarf sample from the \textit{Gaia} Mission}

\author[G. Ourique et al.]{
G. Ourique,$^{1}$
S. O. Kepler,$^{1}$
A. D. Romero,$^{1}$
T. S. Klippel$^{1}$
D. Koester,$^{2}$
\\
$^{1}$Instituto de Física, Universidade Federal do Rio Grande do Sul, 91501-900 Porto-Alegre, RS, Brazil\\
$^{2}$Institut für Theoretische Physik und Astrophysik, Universität Kiel, 24098 Kiel, Germany\\
}

\date{Accepted XXX. Received YYY; in original form ZZZ}

\pubyear{2019}

\begin{document}
\label{firstpage}
\pagerange{\pageref{firstpage}--\pageref{lastpage}}
\maketitle

\begin{abstract}
 Since the \Gaia{} data release 2, several works were published describing a bifurcation in the observed white dwarf colour$-$magnitude diagram for $\GBP{}-\GRP{} > 0$. 
 Some possible explanations in the literature include the existence of a double population with different initial mass function or two distinct populations, one formed by hydrogen$-$ and one formed by helium$-$envelope white dwarfs. We propose instead spectral evolution to explain the bifurcation. From a population synthesis approach, we find that the spectral evolution occurs for effective temperature below ${\simeq}11\,000\,\mathrm{K}$ and masses mainly between $0.64\,\mathrm{M}_\odot$ and $0.74\,\mathrm{M}_\odot$, which correspond to around $16$ per cent of all DA white dwarfs. We also find the \Gaia{} white dwarf colour-magnitude diagram indicates a star formation history that decreases abruptly for objects younger than $1.4\,\mathrm{Gyr}$ and a top-heavy initial mass function for the white dwarf progenitors.
\end{abstract}

\begin{keywords}
white dwarfs --  parallax
\end{keywords}





\section{Introduction}\label{sec:intro}
    The \Gaia{} mission \citep{2016A&A...595A...1G} is already one of the most important surveys for the study of white dwarfs. Since its Data Release 2, it provided photometry on three bands, \Gg{}, \GBP{} and  \GRP{}, and parallax and proper motion of more than $200\,000$ white dwarf candidates~\citep{2019MNRAS.482.4570G}. It is expected that by the end of the \Gaia{} mission, about $400\,000$ white dwarfs should be discovered. In its further releases, low-resolution spectra will be available, allowing the spectral confirmation and classification of white dwarfs~\citep{2007ASPC..372..139J}. All these parameters combined could lead us to uncover the properties of white dwarfs, such as effective temperature, masses, radii, atmospheric composition and ages. Because white dwarfs comprise the endpoint of more than 97\% of the evolution of all stars of the Milky Way, these parameters will allow the mapping of the initial mass function (\IMF{}) and star formation history (\SFH{}) of the solar neighbourhood~\citep[e.g.][]{2019ApJ...878L..11I}.
    
    \citet{2018A&A...616A..10G} described a bifurcation in the colour-magnitude diagram for $\GBP{}-\GRP{} > 0$, defining a high-luminosity and a low-luminosity region. This bifurcation is not expected for a simple population of white dwarfs.
    They argue that the high- and low-luminosity regions coincide with theoretical cooling sequences of $0.6\,\mathrm{M}_\odot$ and $0.8\,\mathrm{M}_\odot$, respectively, considering pure hydrogen atmosphere models from the Montreal Group~\citep{2006AJ....132.1221H, 2006ApJ...651L.137K, 2011A&A...531L..19T,2011ApJ...737...28B}. 
    They note that two separated peaks in the mass distribution are not expected since \citet{2013ApJS..204....5K} did not report any minimum between $0.6\,\mathrm{M}_\odot$ and $0.8\,\mathrm{M}_\odot$ in the Sloan Digital Sky Survey (SDSS) DA mass distribution, only an extended tail for masses above $0.8\,\mathrm{M}_\odot$.
    Thus, they proposed that the bifurcation is better described by two distinct evolutionary sequences of $0.6\,\mathrm{M}_\odot$, one of pure hydrogen and one of pure helium white dwarfs, corresponding to the high-luminosity and the low-luminosity regions, respectively.
    
    Considering that the bifurcation can only be seen for magnitudes $\Gabs{} > 11.5$ and colour $\GBP{}-\GRP{} > 0$ and becomes barely visible for magnitudes $\Gabs{} > 13.5$ and colour $\GBP{}-\GRP{} > 0.5$, in the further sections of this work we call the region with a lower density of white dwarfs inside these limits as the ``\Gaia{} gap''.
    
    Several works were published trying to explain the origin of the \Gaia{} gap.
    \citet{2018ApJ...860L..17E} proposed that a piece-wise linear initial to final mass relation (\IFMR{}) could describe the \Gaia{} gap, since the slope variations of the \IFMR{} induces variations in the number of stars in each mass interval.
    \citet{2018MNRAS.479L.113K} argue that the \citet{2018ApJ...860L..17E} hypothesis could explain the \Gaia{} gap, but not the difference between the binary fraction between the main-sequence ($\sim 50$\%, depending on the spectral type) and the white dwarf population~\citep[26\% according to][]{2016MNRAS.462.2295H}.
    \citet{2018MNRAS.479L.113K} used a population synthesis approach to study a sample of $13\,928$ white dwarfs within $100\,\text{pc}$.
    They estimated that $36$ per cent of the population is composed by helium atmosphere white dwarfs, assuming a Salpeter \IMF{}~\citep{1955ApJ...121..161S}.
    They reported that the presence of a helium atmosphere white dwarf population by itself cannot reproduce the \Gaia{} gap
    and concluded that the gap is a consequence of a bifurcation in the mass distribution due to a double population, one with masses around $0.6\,\mathrm{M}_\odot$, and a second with masses around $0.8\,\mathrm{M}_\odot$.
    They attributed this massive population to mergers.
    Alternatively, core crystallisation during the white dwarf evolution was considered
    as a possible explanation for the \Gaia{} gap. When the white dwarf core crystallises, it releases energy, slowing down the white dwarf cooling. However, \citet{2019Natur.565..202T} concluded that crystallisation is not responsible for the existence of the \Gaia{} gap,
    but produces a separate low luminosity branch in the HR diagram, the so called Q-branch.

    In this work, we investigate the effects of spectral evolution from hydrogen envelope to helium envelope white dwarfs. Spectral evolution in general has been discussed in several works since \citet{1972ApJ...177..723S}. \citet{2011MNRAS.413.2827C, 2012ApJ...753L..16C} developed evolutionary models of hydrogen envelope white dwarfs that undergo spectral evolution, and turn into helium envelope ones. They concluded from a population synthesis approach that spectral evolution is an essential process to be included in the white dwarf cosmochronology, considering that it changes the cooling rate of the white dwarfs. However, it is not known yet how common spectral evolution is, its relation to the white dwarf mass and effective temperature, or what are its consequences in the colour-magnitude diagram for large samples, such as the \Gaia{} sample. \citet{2018ApJ...857...56R} propose that spectral evolution could be driven by convective mixing, which should occur for effective temperatures lower than $13\,000\,\text{K}$ and hydrogen mass between $\log{(\mathrm{M}_\mathrm{H}/\mathrm{M}_{*})}{\sim}10^{-15}$ and $10^{-6}$.

    \citet{2019MNRAS.482..649O} studied the number of spectroscopic DAs and non-DAs from \citet{2016MNRAS.455.3413K} and \citet{2017AJ....153...10M} and
    found that the helium to hydrogen white dwarf number ratio goes from ${\sim}0.075$ for effective temperatures around $30\,000\,\text{K}$ to ${\sim}0.36$ for effective temperatures around $11\,000\,\text{K}$ and becomes nearly constant for colder temperatures.
    From this results they conclude that most non-DAs are the result of the spectral evolution, since the very late thermal pulse, which produces helium envelope white dwarfs, is predicted not to occur for more than $0.16-0.25$ of all white dwarfs~\citep{2001Ap&SS.275....1B,2006MNRAS.371..263L}.
    \citet{2019MNRAS.482..649O} also calculated the mass distribution of spectroscopic white dwarfs classified as DA, DB and DC. They found that the DC mass distribution is centred around $0.7\,\mathrm{M}_\odot$, while the DA mass distributions present a decrease in number in the same region. They concluded that this decrease in the DA mass distribution is an indicator that a fraction of DAs with masses around $0.7\,\mathrm{M}_\odot$ undergo spectral evolution, turning into DCs. It is important to notice that \citet{2019MNRAS.482..649O} used pure-helium model to estimate parameters for non-DAs, which should lead to an overestimation in their  masses~\citep{2019ApJ...876...67B,2019A&A...623A.177S}, if their envelopes are contaminated by hydrogen.
    
    \citet{2019ApJ...876...67B} proposed that pure helium atmosphere white dwarf models do not describe non-DA white dwarfs for effective temperatures below $11\,000\,\text{K}$.
    They show that pure helium models results in higher masses than expected for non-DAs below $11\,000\,\text{K}$, and hydrogen-helium mixed model mass determinations agree better with the mean mass for non-DAs, similar to the conclusion by \citet{2019A&A...623A.177S}. Since white dwarfs that undergo spectral evolution should present some residual hydrogen on its atmosphere, these results agrees with the spectral evolution hypothesis.
    
    So far several works used population synthesis approaches to study properties from several white dwarf populations~\citep[e.g.][]{1998ApJ...497..870W, 1999MNRAS.302..173G, 2001MNRAS.328..492T, 2002MNRAS.336..971T, 2004A&A...418...53G, 2012ApJ...753L..16C, 2013ASPC..469..109T, 2016MNRAS.456.3729C, 2016NewAR..72....1G, 2016MNRAS.461.2100T}. In our work, we perform Monte Carlo population synthesis using as input the white dwarf evolutionary sequences, the \SFH{} and the final mass function (\FMF{}). The \FMF{} is a function that describes the number of white dwarfs that should be formed in each mass interval.
    The stellar density in the colour-magnitude diagram obtained for the synthetic stars is compared to the colour-magnitude diagram from the \Gaia{} white dwarf photometric candidates.
    The novelty of our population synthesis computation is the inclusion of spectral evolution, assuming a fast transition turning a fraction of the DAs into non-DAs. To include spectral evolution we define the spectral evolution probability, i.e. a function which describes the fraction of white dwarfs that undergo spectral evolution. We study the dependence of the spectral evolution probability on the white dwarf mass and effective temperature. Our main objective in this work is to investigate the origin of the \Gaia{} gap.
    
    This work is organised as follows. In Section \ref{sec:datadesc} we describe the data sample used in this work. In Section \ref{sec:simulation} we present the simulation details and its input parameters. In Section \ref{sec:results} we show the results of our simulations and in the Section \ref{sec:conclusion} we discuss our results and present our conclusions.

\section{Data Sample}\label{sec:datadesc}
    In this work we use the white dwarf candidates catalogue from  \Gaia{} data release 2 presented in \citet{2019MNRAS.482.4570G}. This catalogue contains $486\,642$ objects.
    
    To remove low-quality objects with low probability to be white dwarfs and to limit all objects to the same magnitude limits, the following cuts are applied by us:
    \begin{eqnarray}
          \textsc{phot\_g\_mean\_flux\_over\_error} &>& 10 \\
          \textsc{phot\_bp\_mean\_flux\_over\_error} &>& 10 \\
          \textsc{phot\_rp\_mean\_flux\_over\_error} &>& 10 \\
          \textsc{parallax\_over\_error} &>& 10 \\
          \textsc{phot\_g\_mean\_mag} &<& 20.8 \\
          \textsc{phot\_bp\_mean\_mag} &<& 20.8 \\
          \textsc{phot\_rp\_mean\_mag} &<& 20.8 \\
          \textsc{test\_Pwd} &>& 0.75
    \end{eqnarray}
    The \textsc{test\_Pwd} is the probability of a given object to be a white dwarf. \citet{2019MNRAS.482.4570G} proposed that the high-confidence white dwarf candidates should have $\textsc{test\_Pwd} > 0.75$. Furthermore, to avoid binary white dwarfs with non-white dwarf companions, we discard objects that satisfy any of these two conditions~\citep{2019MNRAS.486.2169K}:
    \begin{eqnarray}
        \text{If }&& (\GBP-\GRP)\,>=0.5\text{:}\nonumber\\
        &&\Gabs > 11.25 + 2.5\times (\GBP-\GRP)\label{eq:cut01}\\
        \text{If }&& (\GBP-\GRP)\,<0.5\text{:}\nonumber\\
        &&\Gabs > 10 + 5\times (\GBP-\GRP)\label{eq:cut02}
    \end{eqnarray}
    Our final sample is composed of $91\,685$ white dwarf candidates.
    
\begin{figure}
        \includegraphics[width=\linewidth]{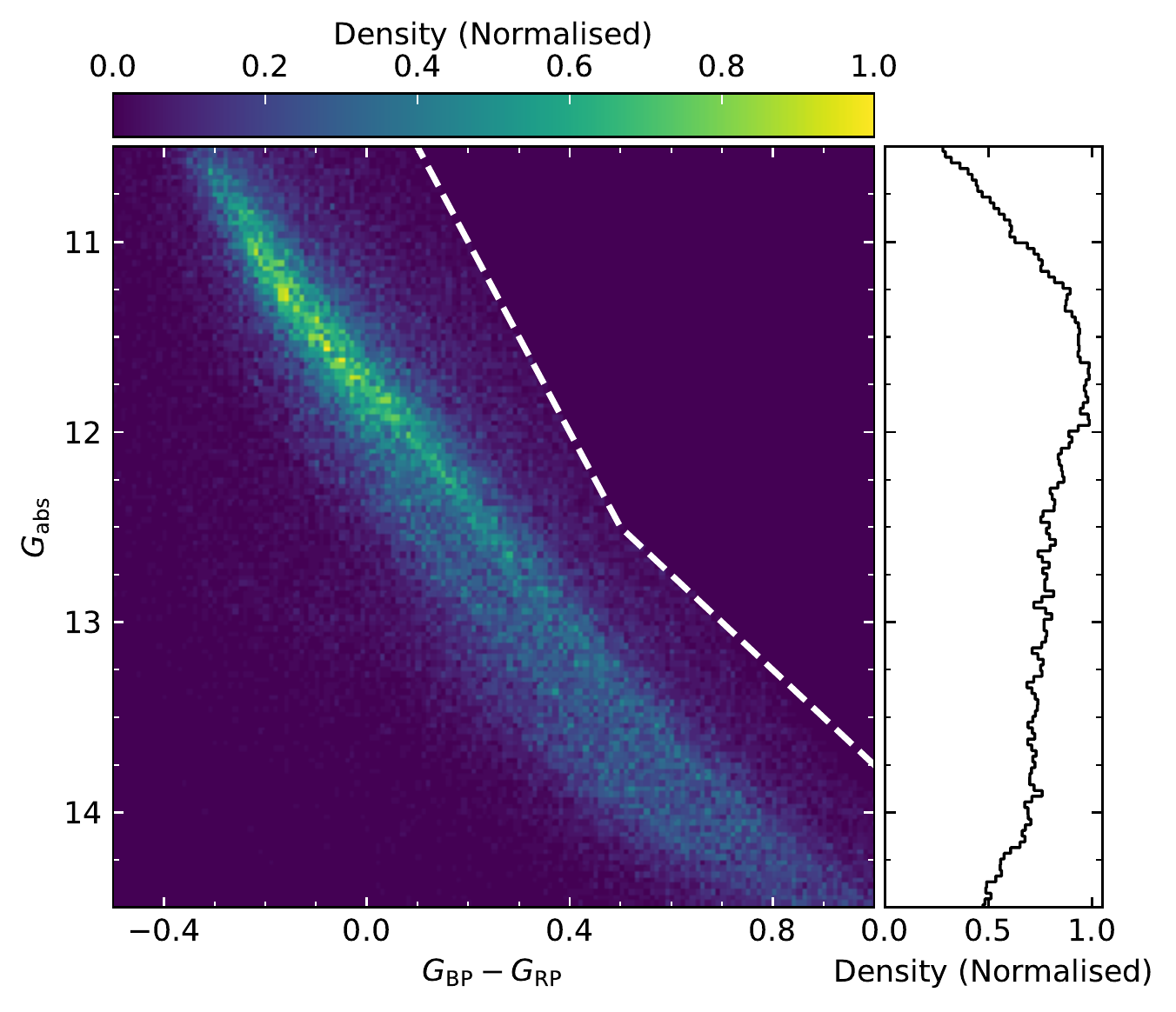}
            \caption{Hess diagram (left panel) from absolute magnitude \Gabs{} and colour $\GBP{}-\GRP{}$ and the luminosity function (right panel) from the absolute magnitude \Gabs{}. In the left panel, the region limited by equations \ref{eq:cut01} and \ref{eq:cut02} are marked as white dashed lines. The region above the white dashed lines is dominated by binaries.}
        \label{fig:Gaia_Hess_LF_NOCUT}
    \end{figure}

    Due to the size of our sample, the stellar density in the colour-magnitude diagram (Hess diagram) analysis becomes a more convenient and precise tool for the analysis, as compared to the scattered stellar colour-magnitude diagram. In Figure \ref{fig:Gaia_Hess_LF_NOCUT} we present the Hess diagram and the luminosity function of our sample. We also marked in the Hess diagram the limits described by equations \ref{eq:cut01} and \ref{eq:cut02}~(dashed white line).
    
\section{Simulation}\label{sec:simulation}
    The main structure of our simulations is based on five components: Evolutionary Sequences, Final Mass Function, Star Formation History, Spatial Distribution and Spectral Evolution. We describe each component below;
    
    \subsection{Evolutionary Sequences}\label{susec:EvolTrack}
    
        In this work we used the hydrogen envelope white dwarf cooling sequences with solar metallicity presented by \citet{2012MNRAS.420.1462R, 2013ApJ...779...58R}. From these sequences, we obtain the stellar mass, $\log{(g)}$, radius and effective temperature as a function of total age.
        We consider the evolutionary sequences with thick hydrogen envelopes, those with the largest amount of hydrogen as predicted by stellar evolution~\citep{2019MNRAS.484.2711R}. The cooling sequences for non-DA white dwarfs are computed by removing the hydrogen layer from a hydrogen atmosphere model at the pre-white dwarf phase, and evolving those models in the cooling curve.
        
        Using a synthetic colour grid~\citep{2010MmSAI..81..921K, 2019MNRAS.486.2169K, 2019A&A...628A.102K} we propagate our models to the colour-magnitude space, using the values of effective temperature and $\log{(g)}$.

    \subsection{Final Mass Function}\label{susec:FMF}

       To perform population synthesis of white dwarfs it is necessary to use a combination of the \IMF{}~\citep[e.g.][]{1955ApJ...121..161S,2001MNRAS.322..231K,2003PASP..115..763C} and the \IFMR{}~\citep[e.g.][]{2005MNRAS.361.1131F,2008ApJ...676..594K,2015MNRAS.450.3708R,2018ApJ...860L..17E} to describe the number of white dwarfs present in each mass interval. We define the function that describes the number of white dwarfs generated in a mass interval as the Final Mass Function~(\FMF{}). Since it would not be possible to directly separate the effect of the \IMF{} and the \IFMR{}~\citep{2019ApJ...878L..11I}, we followed a more convenient approach, that is to define the \FMF{} for our simulations and, from that, extract information on the \IMF{} and the \IFMR{}.

        \subsection{Star Formation History}\label{susec:SFH}
        
        The Star Formation History~(\SFH{}) describes the number of white dwarfs progenitors that are generated in a time interval.
        
        The main-sequence plus giant-branch lifetime of the white dwarf progenitors is strongly dependent on the white dwarf mass.
         For instance, a white dwarf with a stellar mass of $0.85\,\mathrm{M}_\odot$ and effective temperature of $30\,000\,\mathrm{K}$ has a total age of ${\sim}0.2\,\mathrm{Gyr}$, while for a stellar mass of $0.50\,\mathrm{M}_\odot$, the total age at the same effective temperature is ${\sim}13.0\,\mathrm{Gyr}$, for the same metallicity~\citep{2015MNRAS.450.3708R}.
        
        \subsection{Spatial Distribution}\label{susec:SD}
        
    The spatial distribution describes the density of stars that are generated according to their position relative to the Sun.
    Equation (\ref{eq:SD}) describes the assumed spatial distribution, $\rho(r,\theta,\phi)$, where $r$ is the distance, $\theta$ is the Galactic latitude $b$ and $\phi$ is the Galactic longitude $l$.
    \begin{equation}\label{eq:SD}
            \rho(r,\theta,\phi) = e^{\displaystyle -\frac{r\sin{\theta}}{z_v}}
    \end{equation} For convenience we define $z \equiv r|\sin{\theta}|$.
    The $z_v$ parameter represents the combination of the Galactic disk height scale, $z_0$, and the spatial completeness of our sample.

        \subsection{Spectral Evolution}\label{subse:MP}

       Several processes could change the composition of the outer layers of a white dwarf during its evolution~\citep[see][for more detail on these processes]{1987fbs..conf..319F}. In this work, we only consider processes that turn DAs into non-DAs during a simple evolution,
       without external contributions.
       During the cooling, hydrogen envelope white dwarfs reach effective temperatures where the hydrogen layer becomes convective. Depending on the hydrogen layer thickness, it could mix with the underlying helium layer~\citep{1976A&A....52..415K, 1977A&A....61..415V, 1979A&A....74..161D, 1997ApJS..108..339B,2001ApJS..133..413B,2012ApJ...753L..16C}, turning a DA into a non-DA. This process is known as convective mixing.  \citet{2008MNRAS.385..430C,2009MNRAS.396.1709C,2012MNRAS.420.1462R} estimated that the distribution of the hydrogen layer mass contains a fraction between $10^{-4}\,\mathrm{M}_\mathrm{H}/\mathrm{M}_*$, which is the maximum hydrogen layer mass as predicted by simple stellar evolution for a $0.6\,\mathrm{M}_\odot$ white dwarf, and $10^{-10}\,\mathrm{M}_\mathrm{H}/\mathrm{M}_*$.
       This wide hydrogen layer mass distribution indicates that some white dwarfs have layers thin enough to undergo spectral evolution by convective mixing at low temperatures. Diffusion is also a source of spectral evolution, but its effects are mainly important above $T_\mathrm{eff}\simeq 23\,000\,\mathrm{K}$, i.e., above the observed Gaia gap, and therefore it will not be included in our simulation. 
       
      \citet{2018ApJ...857...56R} stated that convective mixing should occur for effective temperatures below $13\,000\,\text{K}$. \citet{1971AcA....21..417P, 2010ApJ...717..183R} and \citet{2012MNRAS.420.1462R} reported, based on evolutionary model calculation, that massive white dwarfs present thinner hydrogen layers when compared to low mass white dwarfs, which makes mixing more probable~\citep{2018ApJ...857...56R,2019MNRAS.482..649O}. Finally, \citet{2008ApJ...672.1144T} and \citet{2010ApJ...712.1345T} estimated the presence of ${\sim}85$ per cent thick hydrogen layers white dwarfs.
        
        The existence of DBAs and DABs, i.e., white dwarfs with helium and hydrogen lines on its spectra, is a evidence that mixing is not necessarily instantaneous~\citep{1997ApJS..108..339B, 2011MNRAS.413.2827C, 2018ApJ...857...56R,2019MNRAS.482..649O}. However, since we do not know the duration of these events, our approximation is that the mixing duration is much smaller than the evolutionary time scale. In our simulations we assume that the mixing is very fast in comparison to the evolutionary timescales, changing the evolution path of the hydrogen envelope white dwarf into a helium envelope white dwarf.
        
        To include the spectral evolution in our simulations, we defined a spectral evolution probability (\SEP{}).
        The \SEP{} determines the probability for a hydrogen envelope white dwarf to turn into a helium envelope white dwarf, as a function of the white dwarf mass.

        \subsection{Population Synthesis}
       
\begin{figure*}
        \includegraphics[width=\linewidth]{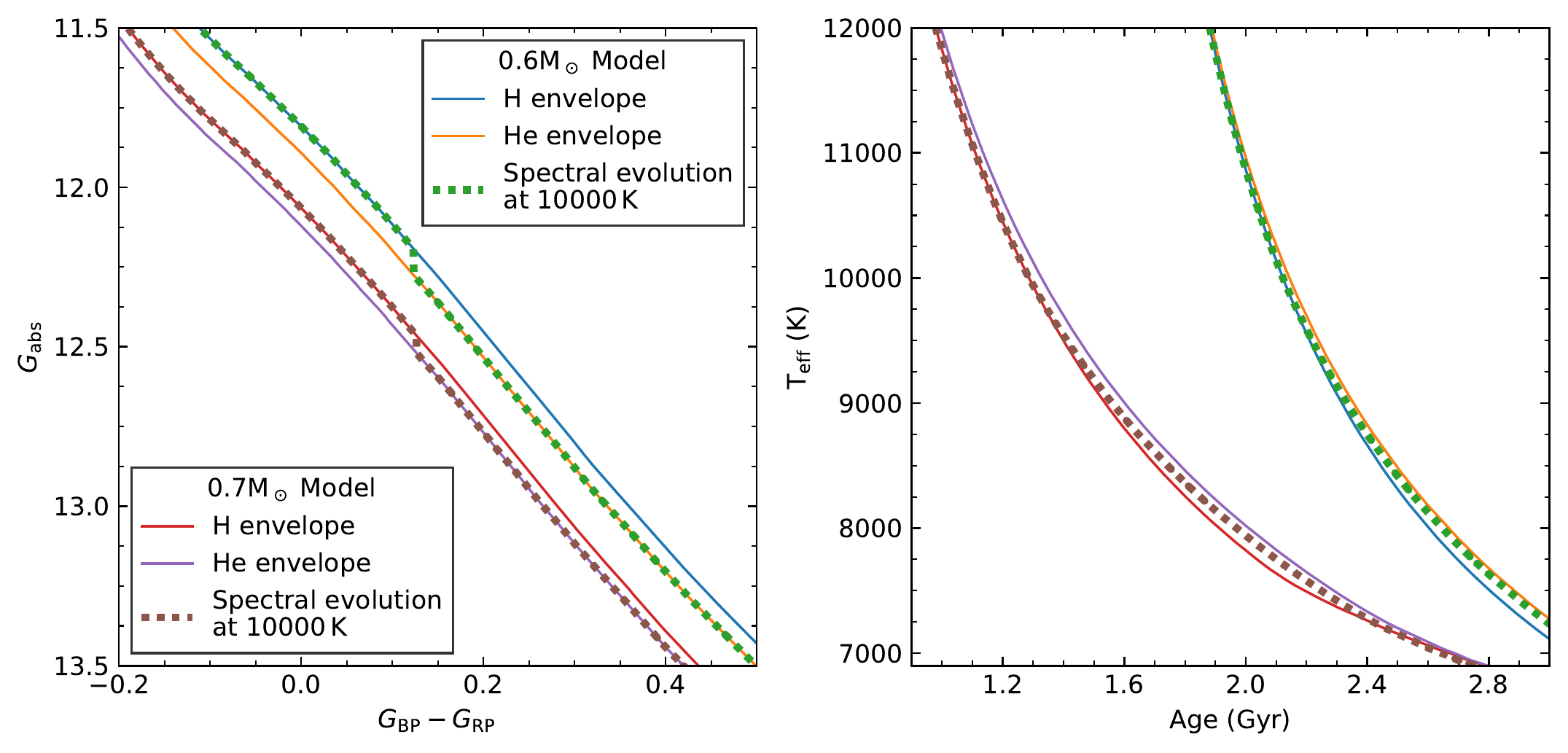}
            \caption{Evolutionary white dwarf cooling sequences.
            The blue and the red solid line indicates the evolutionary path for hydrogen envelope model with masses $0.6\,\mathrm{M}_\odot$ and $0.7\,\mathrm{M}_\odot$, respectively.
            The orange and the lilac solid line indicates the evolutionary path for a helium envelope model with masses $0.6\,\mathrm{M}_\odot$ and $0.7\,\mathrm{M}_\odot$, respectively.
            The green and the brown dashed line indicates the path of a hydrogen envelope model that turns into helium envelope model at $10\,000\,\text{K}$ due to instantaneous mixing assumed.
            The left panel represents the evolutionary path in the colour-magnitude diagram using the absolute magnitude \Gabs{} and the colour $\GBP{}-\GRP{}$.
            The right panel indicates the evolution of the effective temperature as a function of the white dwarf total age.}
        \label{fig:MixedTrack}
        \end{figure*}

        The synthetic population is generated by generating one million of synthetic white dwarfs with hydrogen or helium envelope with random mass, which follows the given \FMF{}, and a random age, which follows the given \SFH{}.
        A fraction of the hydrogen envelope synthetic white dwarfs undergo the spectral evolution according to their white dwarf mass and the \SEP{}.
        The spectral evolution proposed in this simulation assumes that when the synthetic hydrogen envelope white dwarf reaches a given effective temperature, it start to follows a helium envelope cooling sequence of the same white dwarf mass. In our models,
        this transition is instantaneous and the only modification that is applied to the helium envelope cooling sequence is in the white dwarf total lifetime.
        The lifetime of the star undergoing spectral evolution, in the helium envelope cooling sequence
        at the effective temperature where the spectral evolution occurs,
        is defined as the same of the hydrogen envelope cooling sequence at that effective temperature.
        The lifetime only changes from then on.
        For each mass, we have a model with a hydrogen envelope and other with a helium envelope
        in Figure~\ref{fig:MixedTrack}
        (solid lines).
        Also, we 
        include one representative model that starts as hydrogen envelope and then undergoes convective mixing at $10\,000\,\text{K}$ (dashed lines).
        The left panel represents the evolutionary path along the colour-magnitude diagram and the right panel represents the effective temperature as function of the white dwarf total age.
        
        For each generated synthetic white dwarf, a random uncertainty is added to the magnitudes following a normal distribution similar to the uncertainty in the \Gaia{} data.
        
        \subsection{Synthetic White Dwarf Contribution}
        
        Since brighter white dwarfs have a larger observable volume than fainter ones, it is necessary to weight each synthetic white dwarf with its observable volume. Equation (\ref{eq:wGabs}) describes the weight associated with the observable volume and spatial distribution, $w_{\Gg{}}(\Gabs{})$,
        \begin{equation}\label{eq:wGabs}
        w_{\Gg{}}(\Gabs{}) = \displaystyle \int\limits^{2\pi}_{0}\int\limits_{-\pi/2}^{\pi/2}\int\limits_{d_\mathrm{min}}^{d_\mathrm{max}} \rho(r,\theta,\phi) r^2 \cos{(\theta)}\, dr d\theta d\phi
        \end{equation}
        where $\rho(r,\theta,\phi)$ is the density described by equation (\ref{eq:SD}). The radial integral limits, $d_\mathrm{min}$ and $d_\mathrm{max}$ are the minimal and maximal distances that the object can be observed within magnitudes \Gmin{} and \Gmax{}, respectively.
        The parameters $d_\mathrm{min}$ and $d_\mathrm{max}$ are described by Equations (\ref{eq:wdmin}) and (\ref{eq:wdmax}), respectively.
        \begin{equation}\label{eq:wdmin}
        \small  d_\mathrm{min}=10^{\displaystyle \left[1+\scalebox{0.9}{$\frac{\left(\Gmin{}-\Gabs{}\right)}{5}$}\right]}
        \end{equation}
        \begin{equation}\label{eq:wdmax}
         \small  d_\mathrm{max}=10^{\displaystyle \left[1+\scalebox{0.9}{$\frac{\left(\Gmax{}-\Gabs{}\right)}{5}$}\right]}
        \end{equation}
        
        The contribution of each synthetic white dwarf to the Hess Diagram and the luminosity function is given by $w_{\Gg{}}(\Gabs{})$.
        
    \subsection{Uncertainty}
    
    To ensure a trustworthy data analysis, we resampled the magnitudes and parallax over the photometric astrometric and uncertainties from the \Gaia{} mission data, $1\,000$ times,
    assuming a normal distribution of the astrometric and photometric data,
    and calculate the Hess diagram and the luminosity function. For each bin in the Hess diagram and in the luminosity function, we have $1\,000$ density determinations. From the median of these $1\,000$ determinations we define our expected density and from the difference of the $16\%$ and $84\%$ with the median, the lower, and upper density uncertainties.
 
    Since our main focus in this work is the to explain the existence of the \Gaia{} gap, we limit our analysis to the hydrogen envelope white dwarf models of masses between $0.50\,\mathrm{M}_\odot$ and $0.85\,\mathrm{M}_\odot$, comprising the entire \Gaia{} gap structure and avoiding the binary evolution region.
    
\begin{figure}
        \includegraphics[width=\linewidth]{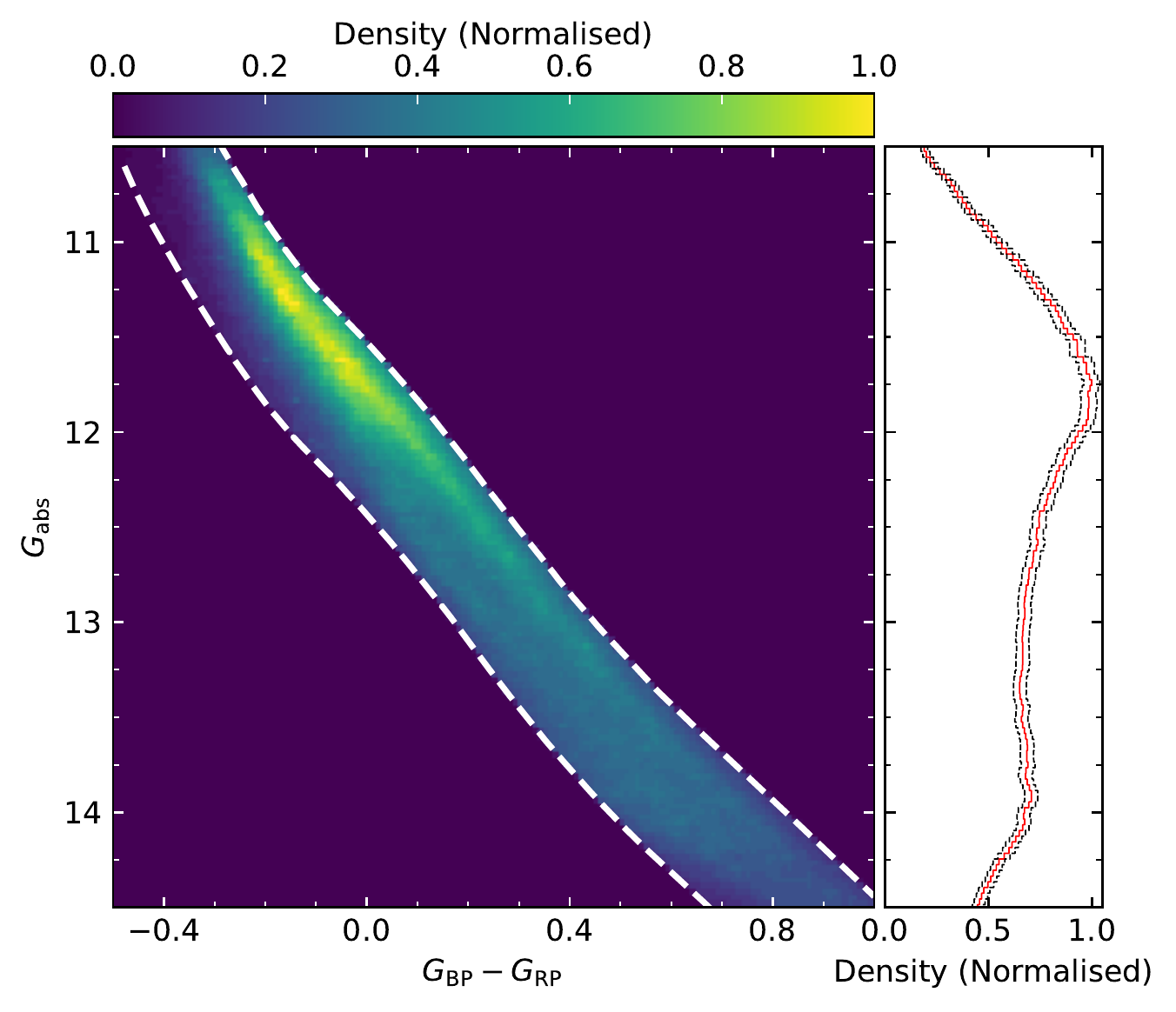}
            \caption{The Hess diagram (left panel) and luminosity function (right panel) of the \Gaia{} white dwarf candidates resampled over their photometric and astrometric uncertainties. The width of the colour bins in the Hess diagram is $0.01\,\mathrm{mag}$. The width of the magnitude bins in the Hess diagram and in the luminosity function is $0.03\,\mathrm{mag}$. The white dashed lines represent our hydrogen white dwarf models of masses $0.50\,\mathrm{M}_\odot$ (brighter) and $0.85\,\mathrm{M}_\odot$ (fainter).}
        \label{fig:Gaia_Hess_LF}
        \end{figure}

    In the left panel of Figure \ref{fig:Gaia_Hess_LF} we show the Hess diagram of the resampled data, and in the right panel, the luminosity function of the \Gaia{} white dwarf candidates described in Section \ref{sec:datadesc}. The white dashed line on the higher luminosity diagonal of the diagram represents a hydrogen envelope white dwarf model of $0.50\,\mathrm{M}_\odot$, while the white dashed line on the lower luminosity diagonal represents the model of $0.85\,\mathrm{M}_\odot$. 
    
    In Figure \ref{fig:Gaia_Hess_LF} we can see the \Gaia{} gap is centred around $\GBP{}-\GRP{}{\simeq}0.2$ and $\Gabs{}{\simeq}12.5$
    and a double peak in the luminosity function.
    We can also notice that the density tends to zero for blue high luminosity white dwarfs ($\Gabs{} < 11.5$).
     
\begin{figure}
        \includegraphics[width=\linewidth]{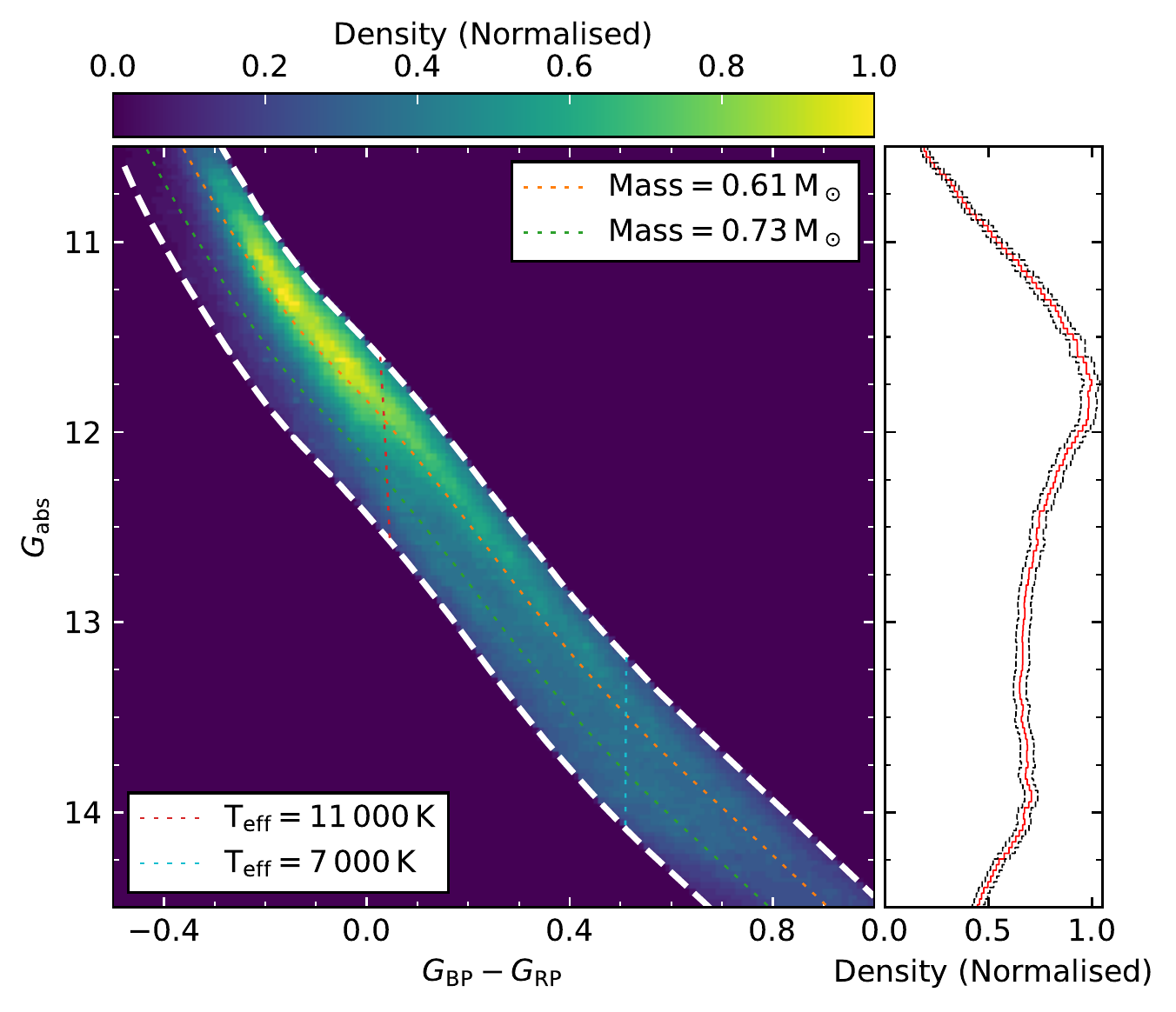}
            \caption{Same data as Figure \ref{fig:Gaia_Hess_LF}. The orange and the green lines represent our hydrogen white dwarf models with $0.61\,\mathrm{M}_\odot$ and $0.73\,\mathrm{M}_\odot$, delimiting the \Gaia{} gap in luminosity. The red and cyan line represents the effective temperature of $11\,000\,\text{K}$ and $7\,000\,\text{K}$, delimiting the \Gaia{} gap in colour.}
        \label{fig:Gaia_Hess_LF_MARK}
        \end{figure}

   To highlight the region of the \Gaia{} gap, we plot in Figure \ref{fig:Gaia_Hess_LF_MARK}
    --- similar to Figure \ref{fig:Gaia_Hess_LF} --- the hydrogen envelope models of $0.61\,\mathrm{M}_\odot$ (orange line) and of $0.73\,\mathrm{M}_\odot$ (green line), and the line of effective temperature $11\,000\,\text{K}$ (red line) and $7\,000\,\text{K}$ (cyan line).
    
\section{Results}\label{sec:results}

    For our population synthesis input parameters, we started with a flat \FMF{} and \SFH{}. We vary these functions in order to minimise the quadratic difference between the Hess diagram from the \Gaia{} data (Figure \ref{fig:Gaia_Hess_LF}) and the Hess diagram from the population synthesis. The same technique was used to estimate the best value for the $z_v$ parameter, which resulted around $110\,\mathrm{pc}$
    due to our restrictions of $10\sigma$ on the parallax and photometric uncertainties.
    
\begin{figure}
        \includegraphics[width=\linewidth]{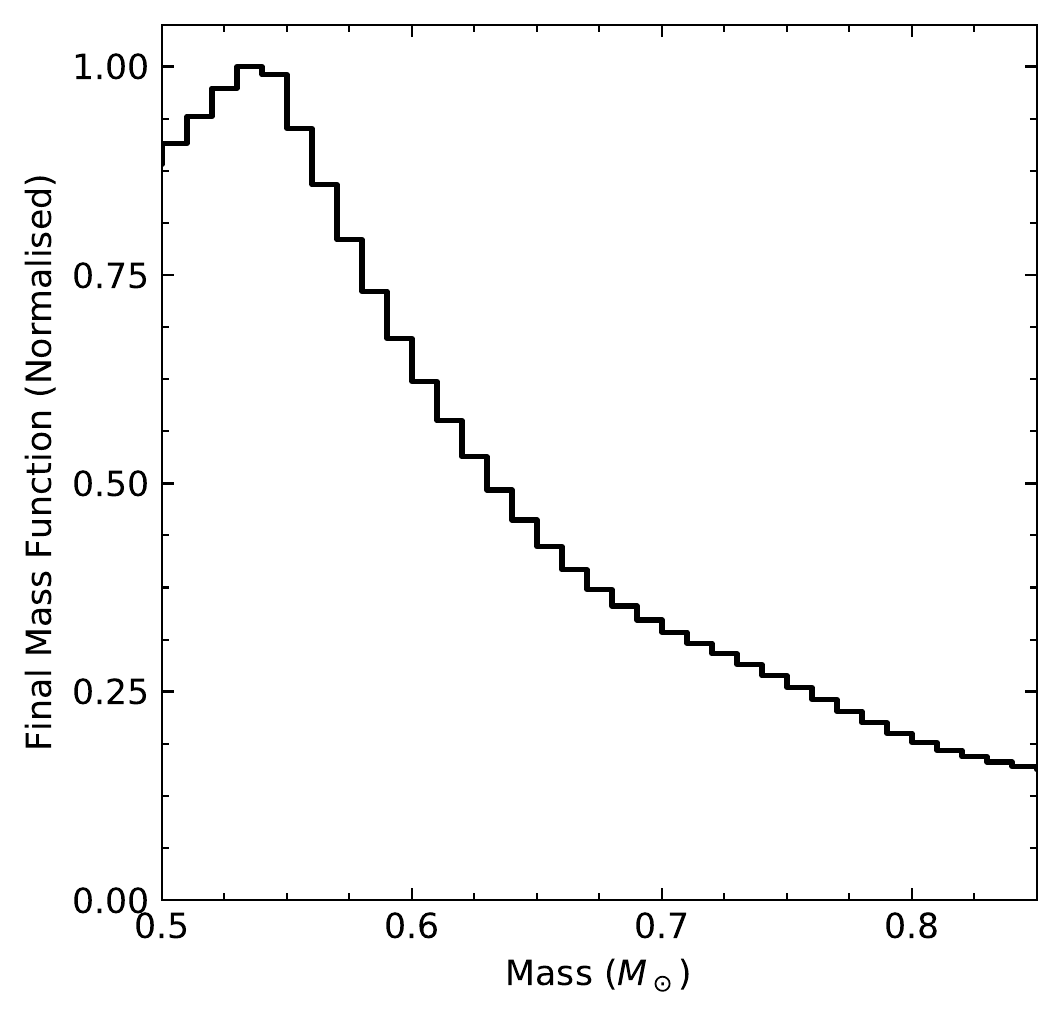}
            \caption{The estimated normalised \FMF{} that minimises the quadratic difference of the Hess diagram of the \Gaia{} mission data and the Hess diagram of our population synthesis. The \FMF{} presents a peak around $0.55\,\mathrm{M}_\odot$, a very low density for masses below the peak and an extended tail for higher masses.}
        \label{fig:Parameter_FMF}
    \end{figure}

    The estimated \FMF{}, presented in Figure \ref{fig:Parameter_FMF}, shows a single peak structure around $0.55\,\mathrm{M}_\odot$ with an extended tail for higher masses.
    
\begin{figure}
        \includegraphics[width=\linewidth]{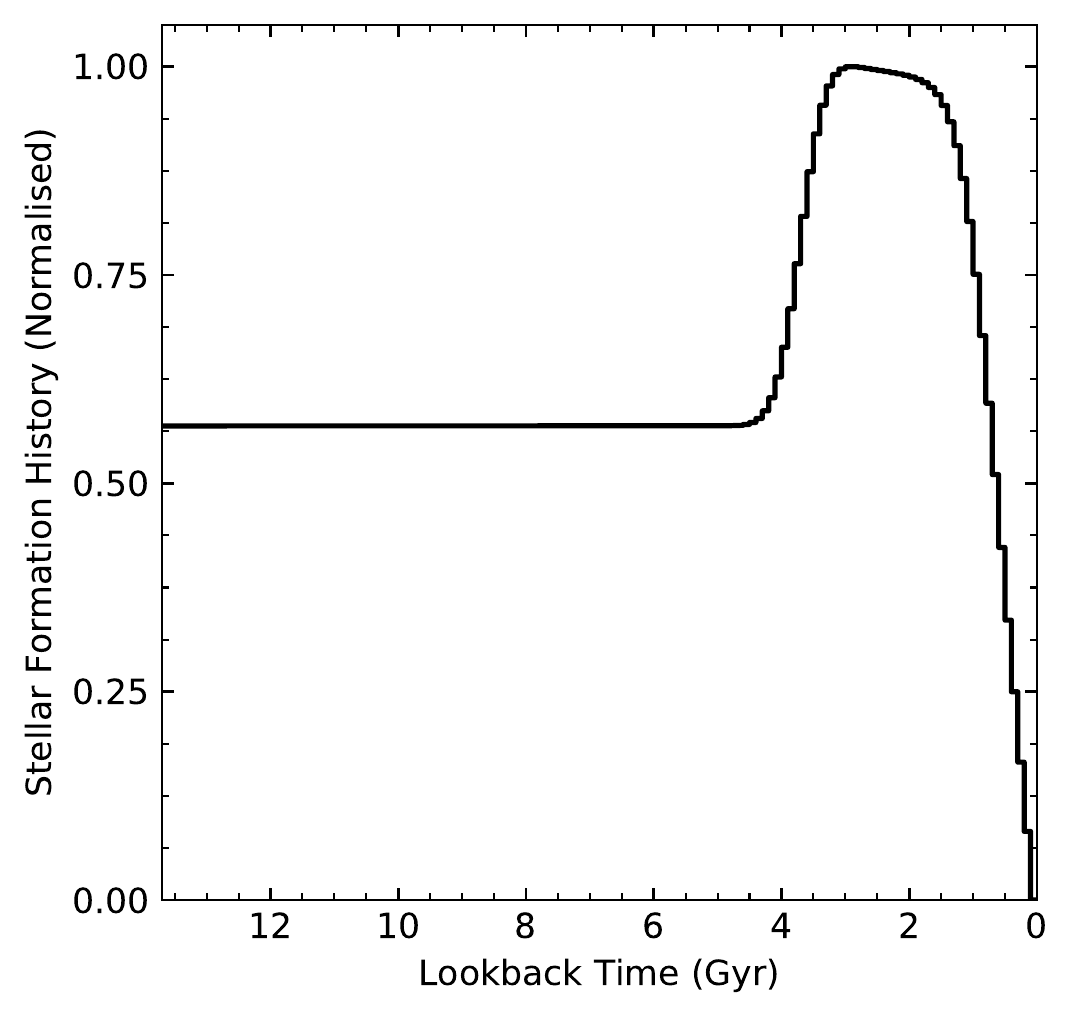}
            \caption{The estimated normalised \SFH{} that minimises the quadratic difference of the Hess diagram of the \Gaia{} mission data and the Hess diagram of the population synthesis. The formation rate between $1.4\,\mathrm{Gyr}$  and $3.5\,\mathrm{Gyr}$ ago is ${\simeq}50$\% higher than the mean \SFH{}.  Also, the \SFH{} presents a very small formation rate for very young stars and a flat formation rate for objects older than $6\,\mathrm{Gyr}$.}
        \label{fig:Parameter_SFH}
    \end{figure}

    The \SFH{}, presented in the Figure \ref{fig:Parameter_SFH}, shows an abrupt decrease in the formation rate for 
    stars younger than $1.4\,\mathrm{Gyr}$. The formation rate for ages between $1.4\,\mathrm{Gyr}$ and $3.5\,\mathrm{Gyr}$ ago shows a ${\simeq}50$ per cent increase over the mean \SFH{}. The flat structure of the \SFH{} for objects older than $6\,\mathrm{Gyr}$ is a consequence of the observed
    $10\sigma$ Gaia data not sampling these cool and faint white dwarfs.
    
\begin{figure}
        \includegraphics[width=\linewidth]{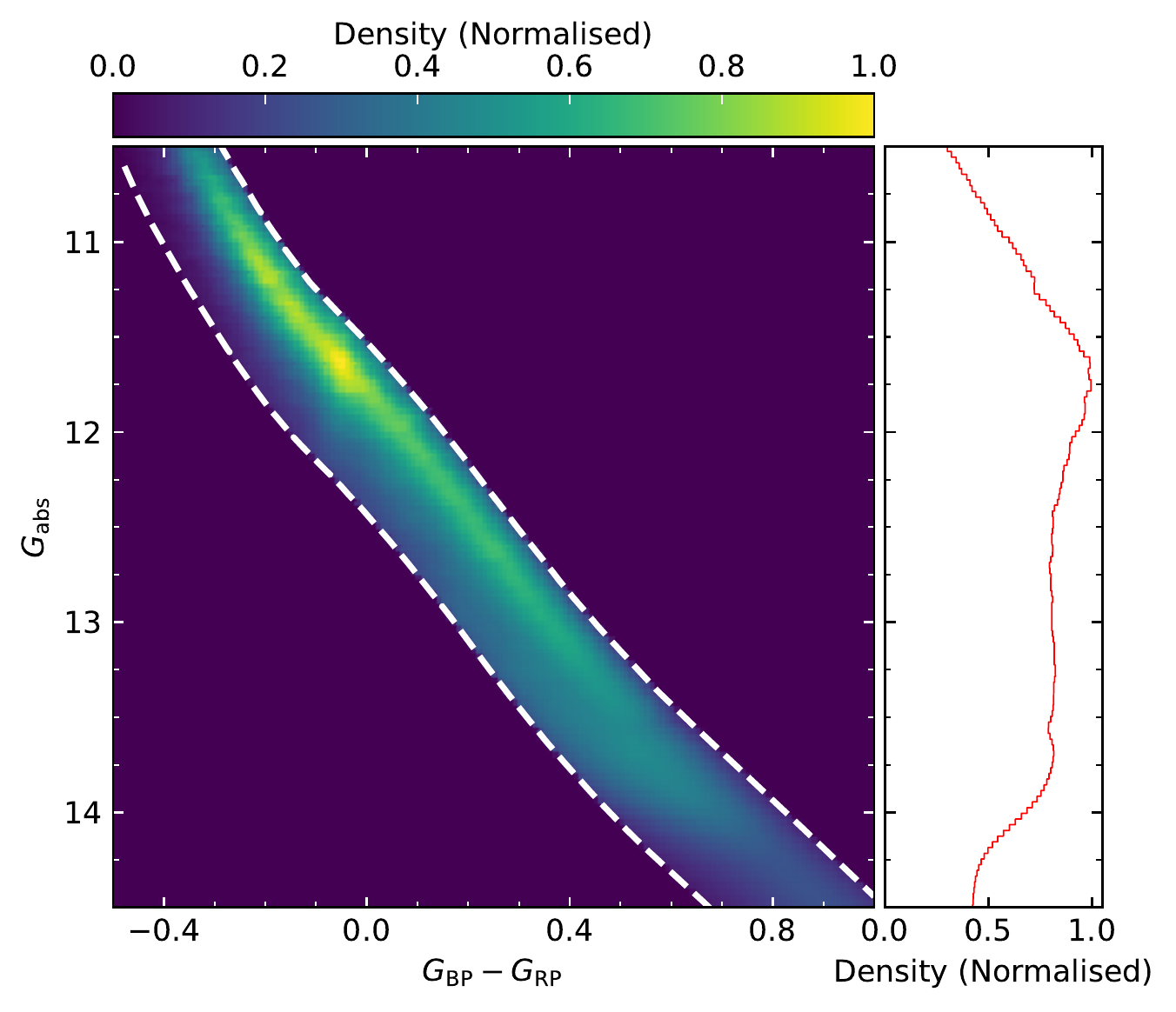}
            \caption{Hess diagram and luminosity function of the synthetic white dwarfs, generated with the \FMF{} from Figure \ref{fig:Parameter_FMF}, \SFH{} from Figure \ref{fig:Parameter_SFH}, $z_v=110\,\mathrm{pc}$ and without spectral evolution. $78$ per cent of the population is composed by hydrogen envelope white dwarfs and the remaining $22$ per cent is composed by helium envelope white dwarfs. The higher luminosity region of the Hess diagram presents a very low density of massive objects, which is a consequence of the small recent star formation rate. The double peak in the luminosity function can be seen in the simulation. The \Gaia{} gap is not reproduced in this simulation, which does not include spectral evolution. The orange and the green lines represent our hydrogen white dwarf models with $0.61\,\mathrm{M}_\odot$ and $0.73\,\mathrm{M}_\odot$. The red and cyan line represents the effective temperature of $11\,000\,\text{K}$ and $7\,000\,\text{K}$. The Hess diagram and the luminosity function are normalised to their maximum value.}
        \label{fig:NoMix_Hess_LF}
    \end{figure}

  Figure \ref{fig:NoMix_Hess_LF} shows the Hess diagram and the luminosity function of the
  best synthetic model, with the \FMF{} and \SFH{} presented in Figures \ref{fig:Parameter_FMF} and \ref{fig:Parameter_SFH}, respectively, with $z_v=110\,\mathrm{pc}$ and without spectral evolution.
  In this population synthesis, we have two populations --- one of hydrogen envelope white dwarfs, which correspond to $78$ per cent of the simulated stars, and the remaining $22$ per cent as helium envelope white dwarfs. These fractions were chosen to agree with the fraction of DAs and non-DAs from the SDSS spectroscopic determinations up to Data Release 14 presented by \citet{2019MNRAS.486.2169K}. Our tests show that the Hess diagram and the luminosity function are very weakly sensitive to the fraction of primordial helium envelope white dwarfs.
  
\begin{figure}
        \includegraphics[width=\linewidth]{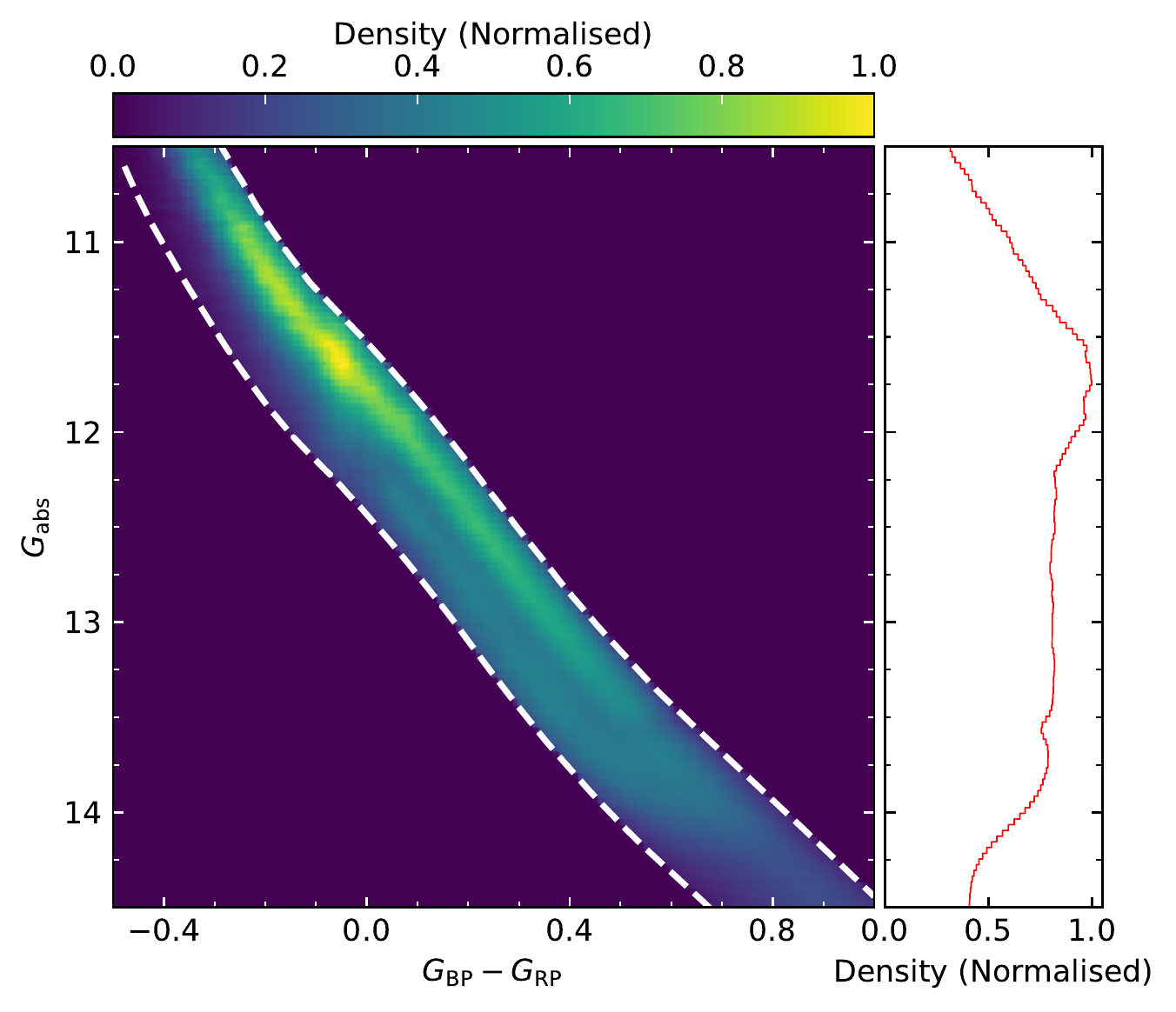}
            \caption{Hess diagram and luminosity function of the synthetic white dwarfs generated with the 
            Final Mass Function (\FMF{}) shown in Figure \ref{fig:Parameter_FMF}, Star Formation Rate (\SFH{}) shown in  Figure \ref{fig:Parameter_SFH} and completeness modulated scale height $z_v=110\,\mathrm{pc}$. $88$ per cent of the population is composed by hydrogen envelope white dwarfs and the remaining $12$ per cent is composed by helium envelope white dwarfs. In this simulation, around $15$ per cent of hydrogen envelope white dwarfs undergo spectral evolution, turning into helium envelope white dwarfs. The features in this Figure are very similar to Figure \ref{fig:NoMix_Hess_LF}, and, in this case, the \Gaia{} gap is reproduced. The orange and the green lines represent our hydrogen white dwarf models with $0.61\,\mathrm{M}_\odot$ and $0.73\,\mathrm{M}_\odot$. The red and cyan line represents the effective temperature of $11\,000\,\text{K}$ and $7\,000\,\text{K}$. The Hess diagram and the luminosity function are normalised to their maximum value.}
        \label{fig:99Mix_Hess_LF}
    \end{figure}

 Figure \ref{fig:99Mix_Hess_LF} shows the result of a very similar simulation than the one presented in Figure \ref{fig:NoMix_Hess_LF}. In this case, we changed the initial 
 fraction of hydrogen and helium envelope white dwarfs to $88$ and $12$ per cent respectively. 
 We included spectral evolution in this simulation with $15$ per cent of DAs turning into non-DAs white dwarfs at an effective temperature of $11\,000\,\text{K}$. The non-DAs coming from spectral evolution represent $12$ percent of the total observable white dwarfs.
 These fractions were chosen to keep the final number of DAs and non-DAs similar to the values of the observed sample~\citep{2019MNRAS.486.2169K}. 
 
    The \SEP{} used to generate the population presented in the Figure \ref{fig:99Mix_Hess_LF} defines that $68$ per cent of the white dwarfs that undergo spectral evolution have masses between $0.67\,\mathrm{M}_\odot$ and $0.74\,\mathrm{M}_\odot$, and the spectral evolution occurs when the white dwarfs reaches $11\,000\,\mathrm{K}$.

\begin{figure}
        \includegraphics[width=\linewidth]{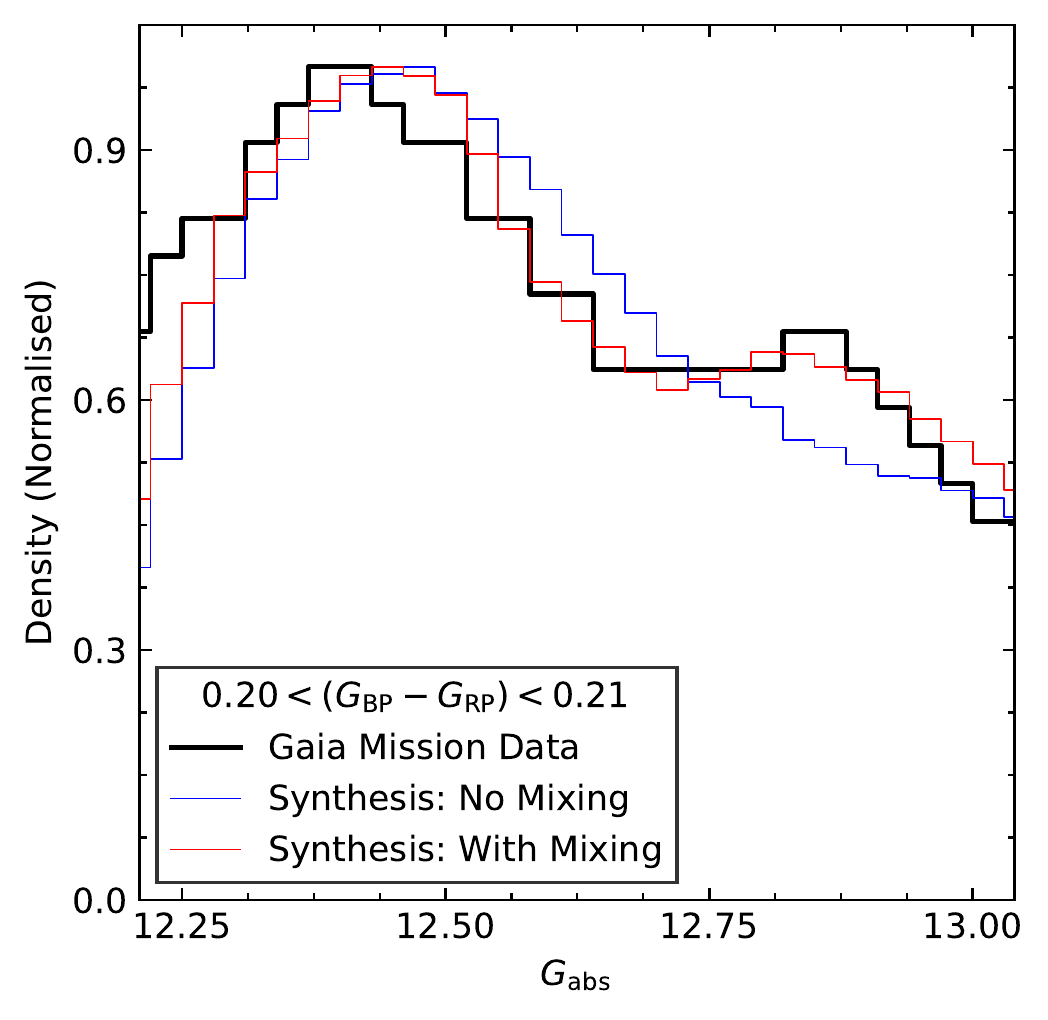}
            \caption{Density distribution of the magnitude $\Gabs{}$ for \Gaia{} mission objects (black solid steps), with no spectral evolution (blue dashed steps) and for spectral evolution (red dashed steps), with colour $\GBP{}-\GRP{}$ between $0.20$ and $0.21$. The \Gaia{} gap occurs between $\Gabs{}=12.35$ and $12.8$, with a minimum around $12.7$. We estimated that the \Gaia{} gap corresponds to around $30$ per cent of the density of white dwarfs. The density presented in this Figure is normalised by the maximum density observed in the Hess diagram. The spectral evolution synthesis can reproduce the profile of the \Gaia{} mission distribution, but cannot reproduce the densities accurately. Each distribution in this Figure is normalised to its maximum.}
        \label{fig:Gaia_IsoColour}
    \end{figure}

    In Figure \ref{fig:Gaia_IsoColour} we present the density distribution for the objects in our sample~(black steps) and for our simulation, with~(red steps) and without~(blue steps) considering spectral evolution, for $\GBP{}-\GRP{}$ between $0.20$ and $0.21$. This Figure indicates that the density of white dwarfs decreases by around $30$ per cent in the \Gaia{} gap in comparison to a linear decrease between the two peaks.
    
    While our spectral evolution synthesis shows a similar 
    profile -- with the density decrease around $\Gabs\simeq 12.75$,
    the synthesis without spectral evolution does not show any increase after this magnitude.
    As our synthesis is a simplified modelling of the several existing parameters, not including, for example, the effect of metallicity and the non-instantaneous spectral evolution, we can reproduce the general features of the \Gaia{} data, but we do not match accurately all regions. We determined that the squared differences between the  Hess diagram for the \Gaia{} mission data and the spectral evolution synthesis is $35$ per cent smaller than the synthesis with no spectral evolution.

\section{Discussion and Conclusion}\label{sec:conclusion}

    The Hess diagram for the \Gaia{} mission data presented in Figure \ref{fig:Gaia_Hess_LF} shows a gap starting around $\GBP{}-\GRP{} > 0$. The position of the centre of the \Gaia{} gap corresponds to the evolutionary path of our hydrogen envelope white dwarf sequence of mass around $0.7\,\mathrm{M}_\odot$. However, since a low white dwarf formation rate with masses around $0.7\,\mathrm{M}_\odot$ would be seen in the entire Hess diagram, it could not be the explanation for the \Gaia{} gap, which is only seen for $\GBP{}-\GRP{} > 0$.
    
    We found that, in the Hess diagram of the colour-magnitude diagram, the density tends to zero for blue high luminosity white dwarfs ($\Gabs{} < 11.5$), a region which corresponds to the massive young white dwarfs.
    In our simulation with no spectral evolution, presented in Figure \ref{fig:NoMix_Hess_LF}, we found that this behaviour is a consequence of a star formation rate that tends to zero in the last billion year (see Figure \ref{fig:Parameter_SFH}), removing the presence of young massive white dwarfs in the local neighbourhood.
    Also, we found that the formation rate for ages between $1.4\,\mathrm{Gyr}$ and  $3.4\,\mathrm{Gyr}$ ago is $50$ per cent higher than the mean \SFH{}.
    
    Our best fit model has a completeness corrected scale height of $z_v=110\,\mathrm{pc}$. This value is much smaller than $z_0=300\,\mathrm{pc}$, estimated by \citet{2017ASPC..509..421K}. This is expected since the $z_v$ parameter represents a combination of the Galactic disk scale height where the white dwarfs are immersed, and the completeness of the $10\sigma$ sample from \Gaia{} data release 2 . The mean distance of our sample is only $190\,\mathrm{pc}$.

    Using the Final Mass Function (\FMF{}) that we obtained from our fit, we estimated the Initial Mass Function (\IMF{}). Using the \citet{2018ApJ...866...21C} \IFMR{}, the resulting IMF is a Salpeter-like function with a slope $\alpha=1.07\pm 0.02$, while using the \IFMR{} of \citet{2015MNRAS.450.3708R} with solar metallicity we found that the slope has $\alpha\,=\,1.38\pm 0.01$. Both results indicate a top-heavy \IMF{}.
    
    Our simulations indicate that the small peak in the luminosity function for magnitudes $\Gabs{} {\sim} 14$ is a consequence of the total time that the star takes to turn into a white dwarf and reach this magnitude.
    For magnitudes $\Gabs{} {\sim} 13$ the age decreases towards higher masses, due to the smaller main-sequence lifetime for higher mass progenitors. 
    However, for magnitudes $\Gabs{} {>} 13$ the age goes through an inflexion.
    Higher mass white dwarfs have shorter cooling rates, leading to a longer age to reach fainter magnitudes than less massive white dwarfs. In numbers, the age of white dwarfs of $0.6\,\mathrm{M}_\odot$, $0.7\,\mathrm{M}_\odot$ and $0.8\,\mathrm{M}_\odot$ at $\Gabs{} {\sim} 14$ are around $3.8\,\mathrm{Gyr}$, $3.4\,\mathrm{Gyr}$ and $3.7\,\mathrm{Gyr}$, respectively. The effect of this inflexion in the luminosity function is a pile-up of white dwarfs around $\Gabs{} {>} 13$.
    
    Figure \ref{fig:99Mix_Hess_LF} shows a population synthesis which includes spectral evolution in the simulation. In this case, we recovered the features from the population synthesis without spectral evolution  --- and the \Gaia{} gap is reproduced. The \SEP{} necessary to reproduce the \Gaia{} gap implies that $\simeq 15$ per cent of hydrogen envelope white dwarfs undergo spectral evolution, turning into non-DAs when their effective temperature reaches $\simeq 11\,000\,\text{K}$. Around $68$ per cent of the white dwarfs that undergo spectral evolution have masses between $0.67\,\mathrm{M}_\odot$ and $0.74\,\mathrm{M}_\odot$. The fraction of white dwarf that undergo spectral evolution corresponds to around $12$ per cent of all the observable sample. 
    
    Since our simulations assumes instantaneous spectral evolution, they do not reproduce accurately the density in the Hess diagram for the region which corresponds to effective temperature between $14\,000\,\text{K}$ and $11\,000\,\text{K}$. This is an indication that spectral evolution should start around $14\,000\,\text{K}$, partially mixing the hydrogen and helium envelopes, and reaching complete envelope transition around $11\,000\,\text{K}$. This effective temperature range is the
    range where H convection zone stars, and it is also the same where most DBAs are located, supporting this conclusion~\citep{2016MNRAS.455.3413K,2015MNRAS.446.4078K,2006ApJS..167...40E,2004ApJ...607..426K}.
    
    The dependence of the spectral evolution on the effective temperature is expected since it is directly related to convection episodes during the white dwarf evolution. 
    \citet{2019MNRAS.486.2169K} show that the mass distribution for DBs has a mean of $0.618{\pm}0.004\,\mathrm{M}_\odot$,
    assuming pure Helium atmospheres, with a very small number of DBs at higher masses, in agreement with \citet{2019arXiv190101857G} and \citet{2019MNRAS.482..649O}. 
    This is an indication that the process that results in extremely thin --- or non-existent --- hydrogen layer white dwarfs (probably Late Thermal Pulse, Very Late Thermal Pulse, or merger) is not effective for the progenitors of white dwarfs with masses higher than $0.75\,\mathrm{M}_\odot$.

   Our simulations indicate that no more than around $16$ per cent of the hydrogen envelope white dwarfs that undergo spectral evolution have masses above $0.74\,\mathrm{M}_\odot$.
  
  Our results also indicate that more than $80$ per cent of all observed helium envelope white dwarfs, i.e. coming from spectral evolution or by primordial hydrogen deficiency, are below $12\,000$, 
   where they do not show helium lines because of the low temperature, in agreement with \citet{2019MNRAS.482..649O} results.
   
   The significant difference in
   the squared difference between the  Hess diagram from the \Gaia{} data and the spectral evolution synthesis Hess diagram of $35$ per cent,
   compared to that with no spectral evolution
    lead us to conclude that the \Gaia{} gap seen in the white dwarf distribution is a consequence of the spectral evolution.
    
    Our simulations are in agreement with \citet{2018ApJ...860L..17E} and \citet{2018MNRAS.479L.113K} results, which shows that the white dwarfs that are found in luminosities below the \Gaia{} gap in the colour-magnitude diagram present a higher fraction between non-DAs and DAs.
    



\section*{Acknowledgements}
We thank the referee for the constructive comments and suggestions that improve this manuscript.

GO, ADR and SOK received support from CNPq and PRONEX-FAPERGS/CNPq (Brazil).
This research has made use of NASA Astrophysics Data System. 

This work has made use of data from the European Space Agency (ESA)
mission {\it Gaia} (\url{https://www.cosmos.esa.int/gaia}), processed by
the {\it Gaia} Data Processing and Analysis Consortium (DPAC,
\url{https://www.cosmos.esa.int/web/gaia/dpac/consortium}). Funding
for the DPAC has been provided by national institutions, in particular
the institutions participating in the {\it Gaia} Multilateral Agreement.

\bibliographystyle{mnras}
\bibliography{GaiaGap} 


\bsp	
\label{lastpage}
\end{document}